# A compact unshielded optically-pumped magnetic gradiometer

Hangfei Ye, Chenlu Xu, Min Hu, and Haifeng Dong*
Instrumental Science Opto-Electronic Engineering, Beihang University

## Abstract

Optically-pumped magnetic gradiometers (OPGs) play a crucial role in applications such as magnetic anomaly detection and bio-magnetic measurements. This study classifies current OPGs into four types based on their differential modes: voltage, frequency, optical rotation, and magnetic field differential modes. We introduce the concept of inherent Common-Mode Rejection Ratio (CMRR) and analyze the differences between the inherent CMRR and the measured CMRR, as well as the upper limit of inherent CMRR. We point out that although magnetic field differential method has the potential to increase inherent CMRR by a factor of 1+AF, the difference between the feedback gains is often neglected, which may set the limit of inherent CMRR. We designed and fabricated a compact, unshielded OPG with a specially designed scheme to minimize the distance between the sensing heads and the magnetic source. Measurement results demonstrate a measured CMRR of 1200@1Hz and a sensitivity of approximately $5 \text{ pT/cm}/\sqrt{\text{Hz}}$ from 1 Hz to 100 Hz.

## I. Classification of OPG Based on the Differential Modes

Optically-pumped magnetic gradiometers have significant applications in various fields such as magnetic anomaly detection [1, 2], bio-magnetic measurements[3, 4], and electromagnetic induction imaging[5]. According to how the gradient information is obtained, OPGs can be classified into four types: voltage differential mode, frequency differential mode, optical rotation differential mode, and magnetic field differential mode.

In the voltage differential mode, the voltage outputs of two separate atomic magnetometers with appreciate baseline are subtracted to obtain the magnetic field gradient information [6-12]. This method is used for magnetocardiography (MCG) [8] and magnetoencephalography (MEG) [7, 13]. Typically, two atomic vapor cells are used in this mode, while a single atomic vapor cell can also be used to measure the field gradiometer within the cell [14, 15]. Diffractive optical elements are employed to increase the baseline in the vapor cell [16]. While the CMRR of this type is not high, off-the-shelf atomic magnetometers can be utilized for the voltage differential mode[12], and the scale factors of the two magnetometers can be calibrated using the measured data to increase the CMRR.

In the frequency differential mode, free induction decay is used to obtain the Larmor precession frequencies at two points. By subtracting the two frequencies, the field gradient can be obtained without calibration[17]. As the gyromagnetic ratio is a fundamental physical constant, this method has the potential to achieve a high CMRR. A CMRR larger than 2000 @60Hz is achieves by Lime et. al.[17]. The nonlinear Zeeman effect and frequency shift are two factors that affect the CMRR of the frequency differential mode gradiometer. Furthermore, high power short pulse laser and high precision frequency counter is needed for this mode.

In the optical rotation angle differential mode, the optical rotations are directly subtracted by double-mirror reflection, half-wave plate, or polarization direction reversal induced by left and right circular pumping [18-21]. Gradiometers which utilize such differential method, also known as intrinsic gradiometers, subtract the optical rotation angle before the differential polarimetry. This approach avoids wasting bits in the AD converter. Additionally, the gain of pre-amplifier can also be increased since the gradient signal is usually much smaller than the common-mode field signal. The disadvantage of this method lies in the fact that the scale factors of the two sensing heads cannot be calibrated offline using the measured data.

The last mode is magnetic field differential mode, which employs a closed-loop system to measure the magnetic field at one magnetometer and feed it back to both magnetometers, making the second magnetometer sensitive only to the magnetic field gradient, thereby achieving direct subtraction of the magnetic field [3, 22, 23]. In the feedback process, one can use either a large coil with a uniform field areas covering multiple magnetometers, or separate coils for different magnetometers[22, 24]. The method can also be used in the second-order gradient measurement by compensating the fluctuations and drifts of background field using controlled oscillators and that of the background gradient by feedback coils[25, 26]. The unique advantage of this method is that the CMRR can be independent on the scale factor difference of the two sensing heads.

## II. The inherent CMRR and its measurement

In this part, we analyze the inherent CMRR and measured CMRR of the first three types of optically-pumped magnetic gradiometers. The power spectrum output of an OPG in the frequency domain is,

$$S_{gr}(s) = \sqrt{k_c(s)^2 B_c(s)^2 + k_{gr}(s)^2 B_{gr}(s)^2} \qquad (1)$$

where $k_c(s)$ is the scale factor of the gradiometer to the common field $B_c(s)$, $k_{gr}(s)$ is the scale factor of the gradiometer to the gradient field $B_{gr}(s)$, assuming that $B_c(s)$ and $B_{gr}(s)$ are uncorrelated.

The inherent CMRR can be defined as the absolute value of the ratio between $k_{gr}(s)$ and $k_c(s)$,

$$\text{CMRR}_{in}(s) = \left| \frac{k_{gr}(s)}{k_c(s)} \right| \qquad (2)$$

Currently, the ratio between the spectral densities of the outputs of the total-field magnetometer and the gradiometer is measured as the value of CMRR[3, 6, 17, 21, 27]. However, the measurement result obtained by the above method is not always equal to the inherent CMRR. It depends on the ratio between gradient and the total field of the measurement environment during the calibration.

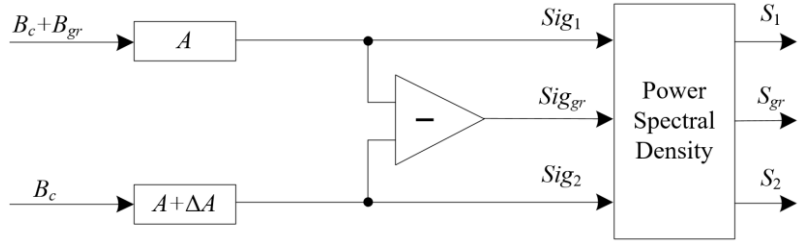

Fig. 1 Scheme for voltage differential mode, frequency differential mode, and optical rotation differential mode OPGs, $B_c$ is the common-mode magnetic field, $B_{gr}$ is the gradient field, $A$ and $A+\Delta A$ are the scale factor of two magnetometers in frequency domain, respectively, and $Sig_{gr}$ is the

**gradient field output signal.**

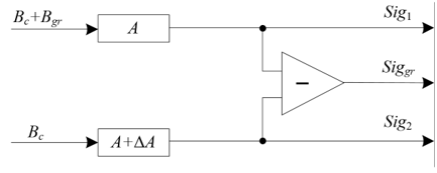

Fig. 1 illustrates the measurement scheme for OPGs in voltage differential mode, frequency differential mode, and optical rotation differential mode. According to the definition of inherent CMRR ($CMRR_{in}$) in equation (2), the inherent CMRR is

$$\mathrm{CMRR}_{in}(s) = \left| \frac{A(s)}{\Delta A(s)} \right| \quad (3)$$

During the calibration of CMRR, $Sig_1(s)$, $Sig_2(s)$ and $Sig_{gr}(s)$ are measured, and $Sig_2(s)/Sig_{gr}(s)$ or $Sig_1(s)/Sig_{gr}(s)$ is usually calculated as the measured CMRR of the OPG.

$$\mathrm{CMRR}_m(s) = \frac{Sig_2(s)}{Sig_{gr}(s)} \approx \sqrt{\frac{1}{\left(\frac{1}{r(s)}\right)^2 + \left(\frac{1}{\mathrm{CMRR}_{in}(s)}\right)^2}} \quad (4)$$

Where $r(s)$ is equal to $B_c(s)/B_{gr}(s)$, and the conditions for approximate equality is that $r(s)$ is much larger than 1, which holds in general cases.

Fig. 2 shows the inherent CMRR and measured CMRR with different $r(s)$. It can be seen that the measured CMRR is always smaller than the inherent CMRR. When the ratio $r(s)$ is smaller than half of the inherent CMRR, the measured CMRR is roughly equal to $r(s)$. The measured CMRR is approximately equal to the inherent CMRR only when $r(s)$ is more than four times the inherent CMRR.

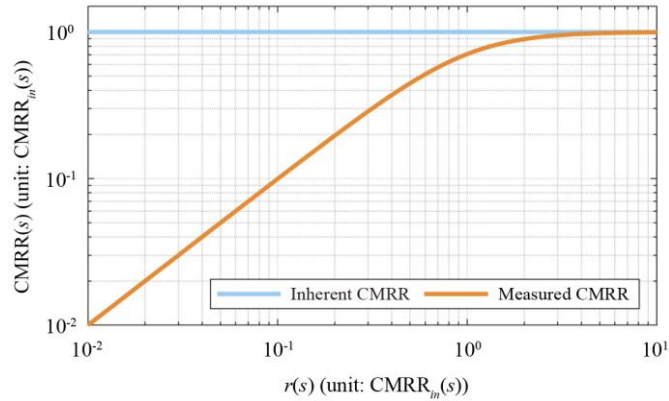

**Fig. 2 Inherent CMRR and measured CMRR with different $r(s)$.**

The scale factor of one of the magnetometers can be adjusted either online or offline to reduce $\Delta A(s)$ in Equation (3), thereby improving $CMRR_{in}(s)$. Fig. 3 illustrates the relationship between $\Delta A(s)$ and the measured CMRR, where $r(s)$ is set to 5000 and $A(s)$ is set to 1000, respectively.

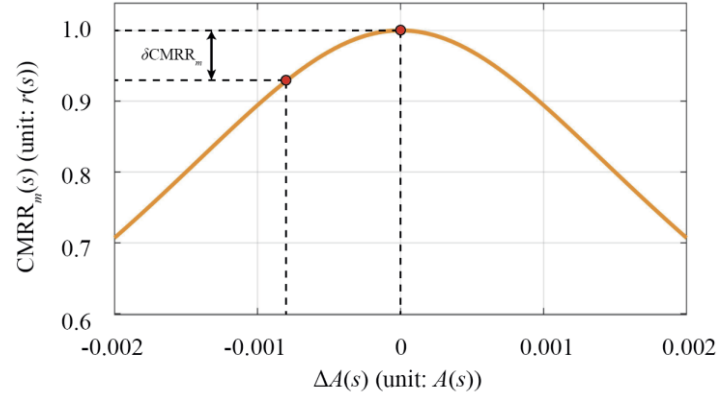

**Fig. 3 A schematic representation of the relationship between the measured CMRR and $\Delta A(s)$.**

From Fig.3, we can see that while the $\Delta A(s)$ can be adjusted by observing CMRRm, the uncertainty of the measured CMRR (i.e., $\delta\text{CMRR}_m(s)$ in Fig. 3) sets a minimum limit on $\Delta A(s)$ and thus a corresponding maximum limit on the inherent CMRR. By analyzing the relation between $\delta\text{CMRR}_m(s)$ and $\Delta A(s)$, we can obtain the upper limit of the inherent CMRR,

$$\text{CMRR}_{in\_\max} = \sqrt{\frac{1}{\left(\frac{1}{r(s)-\delta\text{CMRR}_m(s)}\right)^2 + \left(\frac{1}{r(s)}\right)^2}} \tag{5}$$

Through the analysis above, we arrive at several previously unnoticed points. Firstly, the measured CMRR differs from the inherent CMRR. Secondly, the measured CMRR can be limited by the ratio between the common field and the gradient field. Thirdly, while adjusting the gain difference $\Delta A(s)$ can enhance the inherent CMRR, this approach is constrained by the uncertainty of the measured CMRR.

## III. How to achieve an ultra-high inherent CMRR?

Equation (5) sets a limit for the voltage differential mode, frequency differential mode, and optical rotation differential mode OPGs, assuming that their scale factors are adjustable. For magnetic field differential modes, this limitation can be improved by an additional factor. Ref [3] analyzes the differential output of the OPG under magnetic field differential mode and concludes that the CMRR can theoretically be up to $(1+A(s)F(s))\left|\frac{A(s)}{\Delta A(s)}\right|$, which increases the inherent CMRR by a factor of $1+A(s)F(s)$. However, this has not been demonstrated experimentally. In the analysis of Ref [3], the feedback scale factor is supposed to be exactly the same for the two channels. Here, we reanalyze the system by considering the feedback scale factor difference. The system scheme is shown in Fig. 4.

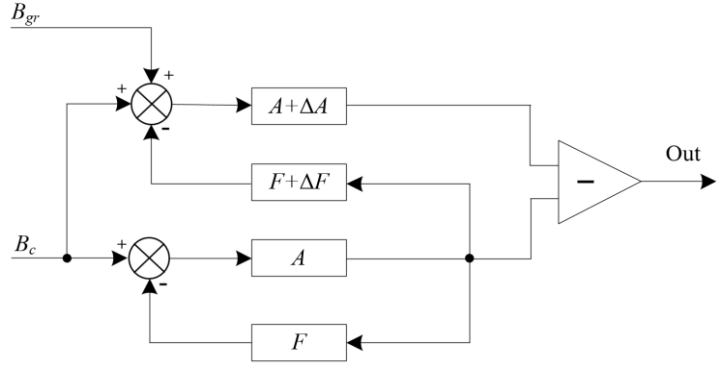

**Fig. 4 Schematic of the closed-loop model after considering the inconsistent coil feedback factor.**

If we consider the gain difference in the feedback, the power spectrum output of the OPG is

$$\text{Out} = \frac{1}{1+AF}\sqrt{\left(-A^2\Delta F - A\Delta A\Delta F + \Delta A\right)^2 B_c^2 + (A^2F + A + AF\Delta A + \Delta A)^2 B_{gr}^2} \qquad (6)$$

where $\Delta F(s)$ is the feedback gain difference of the two channels and $A(s)F(s) \gg 1$. The inherent CMRR is equal to

$$\text{CMRR}_{in} = \left|\frac{A^2F + A + AF\Delta A + \Delta A}{-A^2\Delta F - A\Delta A\Delta F + \Delta A}\right| \approx \begin{cases} \left|\dfrac{F}{\Delta F}\right|, & \Delta F \neq 0; \\ (1+AF)\left|\dfrac{A}{\Delta A}\right|, & \Delta F = 0. \end{cases} \qquad (7)$$

The above equation indicates that, once $\Delta F(s) \neq 0$, the factor limiting the inherent CMRR actually varies from the difference of the forward scale factor difference $\Delta A(s)$ to the feedback scale factor difference $\Delta F(s)$.

In practice, regardless of whether two separate pairs of coils or one large pair of coils are used to provide magnetic feedback, $\Delta F(s)$ cannot be exactly equal to zero. By using one modulated light to simultaneously for both sensing heads, and locking the modulation frequency through feedback, both sensing heads can have exactly the same feedback scale factor. In this way, an ultra-high inherent CMRR of $(1+A(s)F(s))\left|\dfrac{A(s)}{\Delta A(s)}\right|$ may be achieved.

## IV. Design and realization of a compact unshielded OPG

After analyzing the inherent CMRR of four categories, we designed a compact, unshielded OPG using a single amplitude-modulated light. To shorten the distance between the sensing head and the magnetic source, the optical path was optimized so that there were no photodiodes, transimpedance amplifier, or signal lines around the sensing head, as shown in Fig. 5. The light was modulated using an acousto-optic modulator (AOM) and then passed through a circular polarizer, which was fabricated by bonding a linear polarizer and a quarter-wave plate with their optical axes aligned at a 45-degree angle relative to each other.

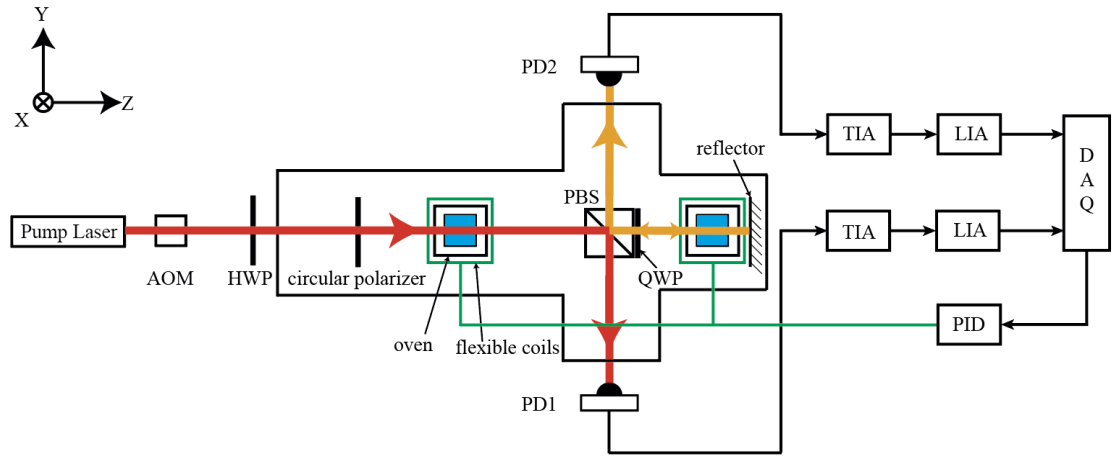

**Fig. 5 Schematic diagram of the experimental setup (AOM: acoustic-optical modulator; HWP: half-wave plate; PBS: polarizing beam splitter; QWP: quarter-wave plate; PD: photodiode; TIA: transimpedance amplifier; LIA: lock-in amplifier; DAQ: data acquisition system; PID: proportional-integral-derivative controller).**

After the light passes through the first atomic vapor cell and the polarized beam splitter (PBS), half of the light was reflected to a photodiode (PD1) for detecting the magnetic field in the first atomic vapor cell. The other half of the light goes into the second atomic vapor cell and is reflected by a mirror behind it. The light is then reflected to PD2 to detect the field in the second vapor cell. The vapor cells had dimensions of 4×4×4 mm$^3$, contained a drop of cesium, and were filled with 650 torr of nitrogen, which served as both a buffer and quenching gas. Each vapor cell was placed in a square boron nitride oven with a side length of 11 mm. Both vapor cells were heated to 85 degrees, resulting in an atomic density of approximately $6 \times 10^{12}$ cm$^{-3}$.

Fig. 6 shows the photos of the OPG after assembly and the main components. Fig. 6 (a) shows the heater fabricated using flexible printed circuit technique and pasted at the bottom of the oven using thermal conductive silicone. The heating line is printed on both side of circuit with a distance of less than 100 μm to eliminate the field generated by the heating current. High-frequency heating at 60 kHz is employed to minimize low-frequency magnetic field interference. Fig. 6 (b) is the coils which control the field around the cells separately. It is also fabricated using flexible printed circuit technique. The coils can be folded into a cube, enveloped around the heater, producing three dimensional magnetic fields around the cells. Fig. 6 (c) is the boron nitride oven with the atomic vapor cell inside.

We designed and fabricated a somos black polymer housing for placing the cells and optical elements using a 3D printing technique. The overall size of the housing was 90×60×18 mm$^3$, with a baseline of 50 mm. The heater, coils, oven, and vapor cell were fixed in the housing with insulating slots around. We used UV glue to secure the PBS and circular polarizer to the housing. Fig. 6 (d) shows the assembly result after removing the housing cap.

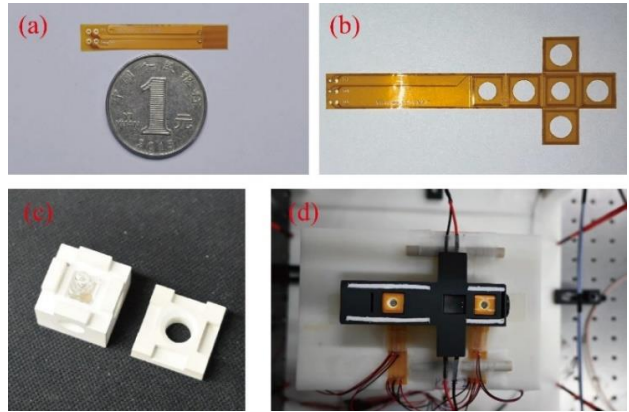

**Fig. 6 Schematic diagram of the core components of the assembled OPG and the overall device: (a) is the flexible heating pad, (b) is the foldable flexible coil, (c) is the boron nitride oven with the atomic vapor cell, and (d) is the assembled OPG unit (without cover).**

After fabrication and assembly, we tested the OPG in a laboratory environment without any shielding. Fig. 7 shows the outputs after lock-in amplifier of the first and the second magnetometers, respectively. The blue line in the graph represents the signals of the first magnetometer, while the orange line represents that of the second magnetometer. The data were recorded continuously for 300 seconds. During the initial 150 seconds, both channels operated in open-loop mode. At 150 seconds, feedback was activated, and the first magnetometer switched to closed-loop. The feedback current also flowed through the coils around the second atomic vapor cell. As most part of the common field was compensated, the output of the second magnetometer directly provided information about the field gradient.

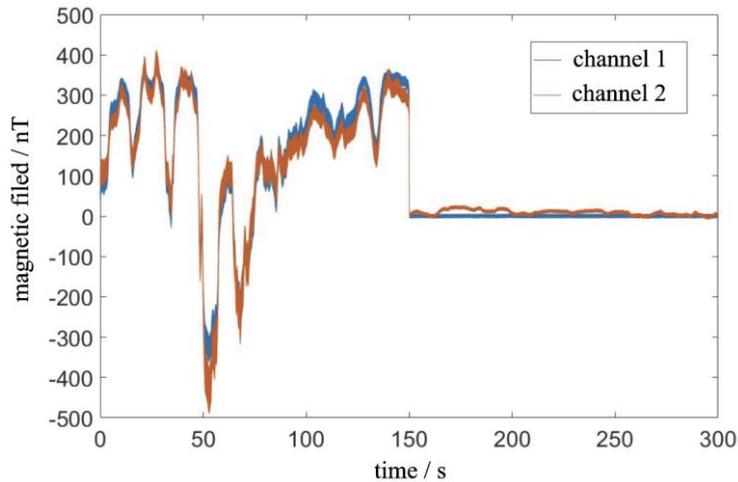

**Fig. 7 Measurement results of both channels in open-loop mode (0-150 s) and closed-loop mode (150-300 s).**

Fig. 8 shows the power spectral density of the OPG in the lab environment without shielding. The blue and red lines represent the power spectral densities of magnetometer one and magnetometer two in the open-loop mode, respectively. The high value of the low frequency field spectral density, reaching up to 1000 nT/$\sqrt{Hz}$, corresponds to environmental field fluctuations. The yellow line represents the spectral density of magnetometer one in the closed-loop mode, showing a significant decrease between 0.1 Hz and 1 Hz compared to the open-loop signal. The purple line corresponds to the output of the second magnetometer, representing the gradient output. The data of these four lines were recorded during working time. The green line, recorded at 2 a.m., represents the same signal as the purple line. It is evident that the gradient fluctuations shown by the purple line and the yellow line at low frequencies are attributed to environmental factors rather than the floor noise of the OPG. Therefore, once the environment gradient field is reduced, the purple line could potentially be lowered to at least the position of yellow line. The CMRR unlimited by the environment field gradient ratio can be calculated by comparing the power spectral densities of the open-loop and the closed-loop signals, which is approximately 2000 @ 0.1 Hz and 1200 @ 1 Hz.

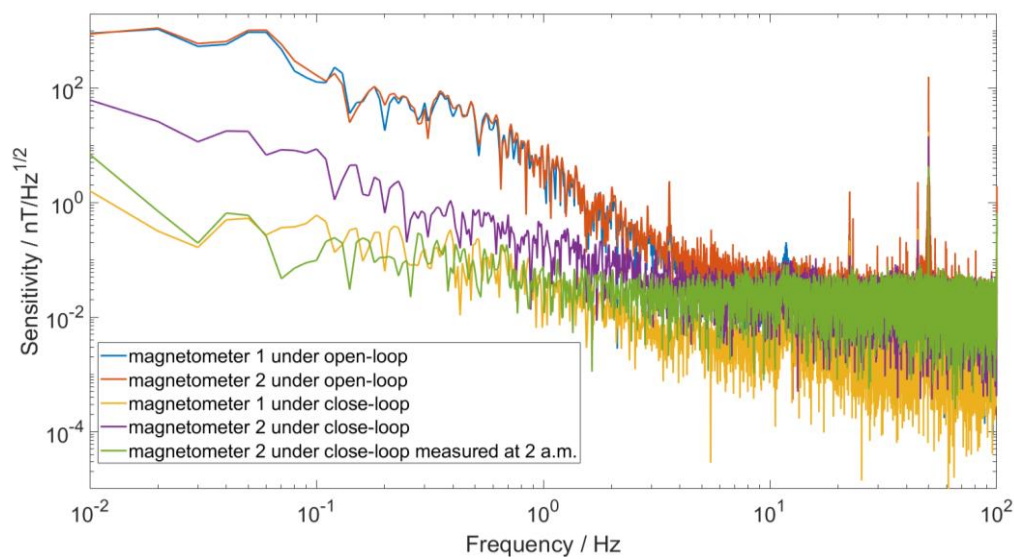

**Fig. 8 The power spectral density of equivalent input magnetic field of the two magnetometers in the OPG, and the light blue line is measured at 2 a.m. while others are all measured in the working time.**

As the OPG is characterized in the lab environment, it is challenging to determine its inherent sensitivity accurately. The test result at 2 a.m. (green line in Fig. 8) provides an upper limit for the sensitivity of the OPG, which is approximately 5 pT/cm/$\sqrt{\text{Hz}}$ within the frequency range of 1 Hz to 100 Hz. It's a pity that

Due to the long response time of the frequency controller of commercial signal function generator, the potential ultra-high inherent CMRR discussed in Part III is not demonstrated in this compact unshielded OPG. The potential ultra-high inherent CMRR discussed in Part III is not fully realized in this compact, unshielded OPG due to environmental noise and frequency feedback delay. We are actively working to address these challenges.

# Conclusion

In this study, we classified various OPGs and defined the inherent CMRR. We analyzed the discrepancy between the inherent CMRR and the measured CMRR as well as the theoretical upper limit of inherent CMRR of the first three type OPGs. We analyze the system of magnetic field differential mode OPG and propose a way to realize the theoretical CMRR enhancement by a factor of $1+A(s)F(s)$ in a real device. A compact OPG with special optical path objecting to shorten the distance between the sensing magnetometer and the magnetic source was designed and characterized in the lab environment without any shielding. A CMRR of 1200@1Hz and a gradient sensitivity of approximately 5 pT/cm/$\sqrt{\text{Hz}}$ at frequency range from

1 Hz to 100 Hz were achieved.